\documentclass[aps,prl, letterpaper,twocolumn]{revtex4}

\usepackage{graphicx}
\usepackage{multirow}

\begin{document}

\title{Distinguishing between Dirac and Majorana neutrinos with two-particle interferometry}



\author{Thomas~D.~Gutierrez}
\address{Nuclear Science Division \\ Ernest Orlando Lawrence Berkeley National Laboratory \\ Berkeley, CA 94720}
\email{tdgutierrez@lbl.gov}



\begin{abstract}
Two-particle interferometry, 
a second-order interference effect, is explored as another possible tool to distinguish between massive Dirac and Majorana neutrinos.  A simple theoretical framework is discussed in the context of several gedanken experiments.  The method can in principle provide both the mass scale and the quantum nature of the neutrino for a certain class of incoherent left-handed source currents.  
\end{abstract}


\maketitle






Two contemporary problems in neutrino physics are determining the absolute mass of the neutrino and discovering if the neutrino is a Dirac or Majorana fermion \cite{nuAPS1}.  The existence of neutrino mass has been established through oscillation experiments such as Super-Kamiokande, SNO, and KamLAND \cite{nuSuper1,nuSNO1, nuKam1}, which have successfully extracted the differences of the squared masses between the energy eigenstates.  Various experimental approaches, such as tritium decay \cite{nuMass1,nuMass2} and cosmological background studies \cite{nuMass3}, are capable of extracting the kinetic mass of the electron neutrino and mass sum of the neutrino energy eigenstates respectively.   While these experiments have been able to put an ever-improving upper limit on the neutrino mass, they provide no information about the neutrino's Majorana or Dirac nature.  
One powerful approach currently used to determine the quantum nature of the neutrino is neutrinoless double beta decay ($\beta\beta(0\nu)$) \cite{nuKay1}.   The decay rate is proportional to the effective mass of the neutrino and only proceeds if the neutrino is a Majorana particle.  Other interesting methods to address these fundamental questions have been explored since the 1950s \cite{nudavis1}, but exploring new ideas may be beneficial.

This Letter investigates another technique, two-particle intensity interferometry, that theoretically provides information about the mass and nature of the neutrino.  This form of interferometry has been used extensively in many areas of physics and has served to cross-pollinate ideas in different sub-fields for over forty years.  It is natural to wonder what role this technology might play in neutrino physics.  

Intensity interferometry  was
originally developed by Robert Hanbury Brown and Richard Twiss (HBT)
as an alternative to Michelson interferometry to measure the angular sizes of stars
in radio astronomy \cite{hbt}.  
The ideas behind intensity interferometry were eventually quantum mechanically applied to photons, rather than classical radio waves, instigating a revolution in modern quantum optics \cite{optglaub1}.
The technology was independently developed in momentum space for final state particles in elementary particle physics 
and is sometimes called femtoscopy in that context \cite{hbtgold60,hbtboal90,hbtbay98,hbtgyu79, hbthei99, hbtSoltzLisa1}.  

The basic observation in two-particle interferometry is pairs of incoherently generated indistinguishable bosons tend to bunch close in phase space while similarly generated fermions tend to anti-bunch.  What ``close'' means exactly depends on the scale and geometry of the problem and in what space one is performing the measurement.  The sensitivity of the effect to the quantum statistics obeyed by measured pairs, in particular the tendency for incoherent fermions to anti-bunch in phase space, is of interest here in an attempt to determine the quantum nature of the neutrino.

A common physical-observable in intensity interferometry is the two-particle correlation function, $C_2$, which is a measure of the degree of independence between two events in some variable of interest, such as momentum, space, or time.  The two-particle correlation function can be written
\begin{equation}
C_2=\frac{P(1,2)}{P(1)P(2)}\sim\frac{{\rm Tr}[\hat{\rho} \hat{a}^{\dagger}_k \hat{a}^{\dagger}_q \hat{a}_k \hat{a}_q]}{{\rm Tr}[\hat{\rho} \hat{a}^{\dagger}_k \hat{a}_k]{\rm Tr}[\hat{\rho} \hat{a}^{\dagger}_q \hat{a}_q]}
\label{c2}
\end{equation}
where $P(1,2)$ represents the joint probability of measuring two events while $P(i)$ represents the individual probabilities of events $i=1,2$ and can be naturally generalized to higher order correlations.  The explicit momentum space form of $C_2$ on the right hand side of Eq.~(\ref{c2}) highlights the basic  physical components of the correlation function.  Tacitly contained in the density matrix, $\hat{\rho}$, when projected as a Wigner function, are the space-time geometry of the source, the source dynamics, and any pairwise interactions.  The quantum statistics of the particles are determined by the (anti-)commutation relations of the creation and annihilation operators, $\hat{a}^{\dagger}$ and $\hat{a}$.
When normalized to the single particle distributions as shown,  $C_2$ is proportional to the relative probability for
a joint two-particle measurement as compared to two single-particle measurements.  If the measurements
are independent, then $C_2$=1.   If the measurements are correlated, $C_2$ deviates from unity.

As Eq.~(\ref{c2}) implies, there are many possible approaches one can use to obtain an explicit expression for the correlation function.  
A particularly simple form for Eq.~(\ref{c2}) that illustrates the essential physics is given by the Koonin-Pratt equation \cite{hbtboal90,hbtkp1}

\begin{equation}
C_2=\int d^3R |\psi(\vec{x}_1,\vec{x}_2)|^2 \rho(\vec{R}).
\label{gsimplec2}
\end{equation}
The equation assumes an incoherent emission of particles from a normalized pair distribution $\rho(\vec{R})$ where $\vec{R}$ is the vector separation between the pairs at the source.  For simplicity, time has been implicitly integrated out of Eq.~(\ref{gsimplec2}).  However, the formalism can be expanded to include correlations in the time domain.  The two-particle wave function, $\psi(\vec{x}_1,\vec{x}_2)$, contains information about the quantum statistics and any pairwise interactions.  Working in natural units ($c=\hbar=1$), if a pair of free identical fermions in any specific triplet spin configuration is considered, the spatial part of the wave function will be antisymmetric upon label exchange (assuming no other quantum numbers are involved) and given by the usual plane wave solution
\begin{equation}
\psi(\vec{x}_1,\vec{x}_2)=\frac{1}{\sqrt{2}}(e^{i\vec{p}_a\cdot \vec{x}_1} e^{i\vec{p}_b \cdot \vec{x}_2}-e^{i\vec{p}_a \cdot \vec{x}_2} e^{i\vec{p}_b \cdot \vec{x}_1}).
\label{psi12}
\end{equation}
One interprets this two-particle wave function as the amplitude for particles emitted at points $\vec{x}_1$ and $\vec{x}_2$ to be measured with momenta $\vec{p}_a$ and $\vec{p}_b$.  
For free particles, $C_2$ is simply related to the cosine transform of the incoherent pairwise source distribution, $\rho(\vec{R})$.

If two identical free fermions are emitted from exactly two point sources separated by $\vec{R}$, Eq.~(\ref{gsimplec2}) can be written 
\begin{equation}
C_2(\vec{Q})=1-\xi\cos(\vec{Q}\cdot\delta\vec{x)}
\label{simplesln}
\end{equation}
where $\vec{Q}=\vec{p}_a-\vec{p}_b$ and $\delta\vec{x}=\vec{x}_1-\vec{x}_2$.   The parameter $\xi=1$ for triplet spin states and $-1$ for singlet states.  If the system is spin-averaged, then $\xi=\frac{1}{2}$.  Notice in the triplet case $C_2(\vec{Q}=0)=0$ and the fermions are anti-correlated if in the same momentum state. 
Because the emission is incoherent, and there are no interactions, the correlations arise only from the quantum statistics obeyed by the particles. The scale of the correlation is set by the source size.  It is instructive to note that for non-identical particles, where the wave function has no particular symmetry, $C_2=1$ for all $\vec{Q}$.   

Let's examine a useful limit of Eq.~(\ref{simplesln}) that will be used later for a series of gedanken experiments.  Consider two point sources of fermions separated by a distance $\vec{R}$ and measured by a pair of distant detectors separated by $\vec{d}$.  The source and detector are a distance $L$ from each other such that $L\gg R \gg d$.  That is, there are well-separated sources far away from a relatively close pair of detectors.  Assume a pair of single-mode fermions. 
In this limit Eq.~(\ref{simplesln}) becomes
\begin{equation}
C_2(d)=1-\xi\cos(\Delta\theta d/\lambda).
\label{bigsource}
\end{equation}
Although the particles here are fermions, this is similar to the original HBT experiment used to measure the angular size of stars.  The correlation function is measured at different detector separations, $d$, for waves of known wavelength, $\lambda$.  From the shape of $C_2(d)$, the angular size, $\Delta\theta$, can be extracted.  

Imagine not knowing {\em a priori} the quantum nature of the particles being measured, but instead knowing some other information such as the angle subtended by the source relative to the detectors.  In that case, using Eq.~(\ref{bigsource}), one would fix the angular size and wavelength but then look at the behavior of $C_2$ as the distance between detectors approached zero to determine the quantum statistics obeyed by the particles of interest.
  
Can two-particle interferometry be applied to neutrinos to determine if they are Dirac of Majorana particles?  Let's examine four variations of a simple gedanken experiment, labeled A through D below,  to answer this question.  A summary of the relevant formulae and the ability of the four cases to resolve the neutrino mass and nature are outlined in Table \ref{table1}.  For simplicity, only one neutrino flavor with one mass eigenstate is considered and oscillations are ignored.  

It will be helpful to remember for the cases below that while Majorana neutrinos are their own antiparticle (the field operators transform to themselves under a charge conjugation operation), the left-handed weak source currents creating them will generate final state particles with a handedness as if they were Dirac fermions \cite{nuKay1}.  
\begin{table}[htdp]
\caption{The two-particle correlation function for Dirac, $C^{\rm Dir}_2(d)$, and Majorana, $C^{\rm Maj}_2(d)$, neutrinos are shown for various situations.  Where $\xi$ alone is quoted, use Eq.~(\ref{bigsource}).  An entry of $C_2=1$ indicates no correlation. The helicity
column notes if detectors are filtering on same, opposite, or averaged final state helicities.  The final rows provide an overview 
of the case-by-case physics capability to determine the neutrino mass or discover the neutrino nature.  Case A: $m=0$, identical sources;  Case B: $m=0$, distinguishable sources; Case C: $m\neq 0$, identical sources; Case D: $m\neq 0$, distinguishable sources.  See the text for a detailed case-by-case discussion.}
\label{table1}
\begin{center}
\begin{tabular}{|c||c|c|c|c|c|}
		
\multicolumn{2}{c}{}  & \multicolumn{4}{c}{Gedanken Cases}	\\
\hline
			&  helicity &A  &  B  &  C  &  D 		 \\
\hline \hline
\multirow{3}{*}{$C^{\rm Dir}_2(d)$}		
& same &	$\xi=1$ & n/a	&	$\xi=1$ & $ C_2=1$	\\ 
& opp &	n/a &	$C_2=1$ &	$C_2=1$  & $C_2=1$ \\ 
& ave &	$\xi=1$ &	$C_2=1$ &	$\xi=1-m^2/E^2$ &  $C_2=1$ \\ 
\hline
\multirow{3}{*}{$C^{\rm Maj}_2(d)$}		
& same &	$\xi=1$ & n/a	&	$\xi=1$ &   $\xi=1$	\\ 
& opp &	n/a &	$C_2=1$ &	$C_2=1$  & $C_2=1$ \\ 
& ave &	$\xi=1$ &	$C_2=1$ &	$\xi=1-m^2/E^2$ &  $\xi=m^2/E^2$ \\ 
\hline
\multicolumn{2}{|c|}{Mass?}  & no & no & yes & yes \\
\multicolumn{2}{|c|}{Nature?}  & no & no & no & yes \\
\hline

\end{tabular}
\end{center}
\label{default}
\end{table}

First, in case A, consider a massless neutrino and a geometric setup like that describing Eq.~(\ref{bigsource}): well-separated sources far away from close detectors.  Imagine two reactors acting as incoherent point sources of indistinguishable particles normally called Dirac antineutrinos. 
Two distant detectors are separated by a distance $d$.  Relative to the detectors, the reactor pair subtends a known angle $\Delta\theta$.  There are two situations:  one where the sources emit right-handed massless Dirac antineutrinos and another where the sources emit right-handed massless Majorana neutrinos.  In this case, measuring $C_2$ cannot distinguish between Dirac and Majorana particles.  The measured correlation function will be equal to that in Eq.~(\ref{bigsource}) with $\xi=1$ and will give the same result for both the Dirac and Majorana cases.  This is because quantum indistinguishability applies equally well for the two situations and the two-particle wave function will be identical in both cases.  Indeed, this is a sanity check because in the massless limit, Dirac and Majorana particles cannot be distinguished based on the Practical Dirac-Majorana Confusion Theorem \cite{nuconf1}.

Next, for case B, consider massless neutrinos with a similar geometric source-detector setup as above except with one of the reactor sources being replaced by a ``small sun''.  That is, there are two sources emitting distinguishable objects: one, an incoherent point source of particles normally called Dirac neutrinos, and another that would again be Dirac antineutrinos.  But similar to case A, not knowing the neutrino nature, there is no way to use $C_2$ to determine if there is one source of Dirac neutrinos and another of Dirac antineutrinos or if there is a pair of sources emitting Majorana neutrinos of opposite handedness.  The correlation function $C_2(d)=1$ for both scenarios.  This is because the two-particle wave function for either has no special symmetry.  That is, it factorizes and the particles are not entangled at the detector.  From Eq.~(\ref{gsimplec2}), if the normalized wave function factorizes, the correlation function becomes unity.

For C and D let's consider the above two cases again but this time give the neutrino a mass that is small compared to its energy.  The presence of mass complicates the situation because chirality (``handedness''), is no longer the same as helicity.  Also, for a realistic Majorana mass term,
like that introduced in the seesaw mechanism, the mass-degenerate four-component Dirac spinor splits into two two-component Majorana spinors.  For the Majorana cases below, we can imagine taking the light doublet, keeping in mind that the value of the mass, $m$, will be different in the Dirac and Majorana cases but both will still be light compared to the mass of other leptons.  

The primary effect of interest is that left-handed weak source currents can now create massive neutrinos and antineutrinos of the ``wrong'' helicity with an amplitude that goes like $m/E$ when $m\ll E$.  This will determine the the probability of measuring an indistinguishable pair in the final state that will be treated separately for the Dirac and Majorana sources.  If the measured fermion pair is indistinguishable, the wavefunction must be antisymmetric.  The probability of this to occur, which will be related to the mass, will determine the strength of the two-particle correlation function.  This is similar in spirit to the considerations in neutral kaon femtoscopy, although using different sources, quantum numbers, and statistics \cite{hbtbekele1}.

With this in mind, consider case C where the source-detector geometry with two reactors is the same as case A.  However, this time each reactor is the source of either Dirac antineutrinos of mixed helicity or Majorana particles of mixed helicity.  For $m\ll E$, the helicity mixture will be mostly $\Lambda=+1$ with some $\Lambda=-1$ in both the Dirac and Majorana cases.  For this exercise, consider ideal detectors that are capable of filtering on the neutrino helicity.  If the detectors filter on identical helicities in the final state, $C_2$ will be Eq.~(\ref{bigsource}) with $\xi=1$, the same as case A.  Particles of opposite helicity are quantum mechanically distinguishable, so if the detectors filter on opposite helicities then $C_2=1$, as in case B.  However, if the detectors helicity-average particles in the final state, the mixed helicty of the source has the effect of introducting a helicity ``contamination'' at the detector and there will be quantum distinguishably for a small fraction of the measurements.  This contamination will have the effect of diluting the correlation function by a factor $\vartheta(m^2/E^2)$; so use Eq.~(\ref{bigsource}) but with $\xi\sim(1-m^2/E^2)$ for both Dirac and Majorana particles.  Again, Dirac and Majorana neutrinos cannot be distinguished, but a careful helicity-averaged measurement of $C_2(d)$ could, in principle, extract the mass by measuring the strength of this weak anti-correlation.

Finally, in case D, revisit the non-identical sources of ``sun-reactor'' geometry of case B, but extend it to the massive neutrino case.  Because of helicity mixing, the quantum distinguishability arguments are similar to C but now there are more combinatorics for the Dirac particles because of the extra lepton quantum number.  Nevertheless, like the massless case, the Dirac particles are always distinguishable at the detector either by helicity or by lepton number.  No matter how one filters on the final state, the Dirac particles are distinguishable so $C_2=1$.  

If the neutrino is a Majorana particle, however, case D will be different.  The reactor source will be emitting primarily Majorana neutrinos with $\Lambda=+1$ with a small component of $\Lambda=-1$.  The sun source will be emitting Majorana neutrinos of the opposite degree of contamination: mostly $\Lambda=-1$ with a small $\Lambda=+1$ mixture.  Here, because the Majorana neutrino is its own antiparticle, all emitted neutrinos are just various helicity states of the same particle.  With a judicious choice of filtering at the detector, one could detect a distinct signal compared to the Dirac case.  For example, if the detectors filter on opposite final state helicity, $C_2(d)=1$ because the particles are distinguishable.  But if the detectors filter on the same helicity, $C_2$ becomes Eq.~(\ref{bigsource}) with $\xi=1$.  If a helicty-averaging is performed in the final state, this introduces contamination (more severe than case C) that will reduce the correlation strength.  The probability of measuring two equal helicity states with open final-state helicity filters scales like $m^2/E^2$ so use Eq.~(\ref{bigsource}) with $\xi\sim m^2/E^2$.  The neutrinos would very nearly be anti-correlated at small $d$, with only a slight deviation given by $\xi$.

Let's entertain some experimental considerations.  The primary concerns are data rate, detector efficiency, and energy resolution.  The above discussion assumed infinite energy resolution to resolve neutrinos of an arbitrary wavelength with no loss of fidelity or smearing.  This assumption, using Heisenberg's Uncertainty Principle, permits infinitely slow counting statistics, allowing quantum mechanically coherent data to arrive over infinitely long time scales.  This is clearly an unrealistic practical assumption.  

The data rates for current experiments such as KamLAND and SNO are about one event per day.  To perform the measurement, even assuming copious statistics, the ability to measure neutrinos of arbitrary energy, and very fine vertex resolution, experiments would require an unphysical energy resolution to see the effect as described.  Conversely, using $\Delta E\Delta t\sim$ (eV)(fs) it can be seen that even with extremely good, but still physical, energy resolutions, eV or keV, an experiment needs to measure neutrino pairs separated by times on the order of femto- to attoseconds -- a rate approaching weak-charge Amperes of neutrinos.  If neutrinos could, in principle, be measured experimentally with such copiousness and efficiency, other methods would mostly likely provide a more straightforward path to revealing the neutrino's currently unknown properties to the same order in $m^2/E^2$.

The femtoscopic limit of Eq.~(\ref{simplesln}) ($L\gg d \gg R$) can also be considered.  In that limit, neutrinos and antineutrinos could be generated from very small sources like those created in a high energy physics collisions.   In order to image femtometer-sized sources, an experiment would construct $C_2$ in momentum space, measuring two or more identified or reconstructed inclusive neutrinos per event with a momentum resolution of roughly MeV.   Finally, the method could be applied as an anti-bunching counting experiment in the time domain, similar to what is done in quantum optics with photons.  This could be performed on a beam of neutrinos and/or antineutrinos, mirroring cases A-D above.  High flux neutrino-antineutrino beams, like those expected from muon colliders, and exceptional detection time resolution would be required.

Based on the gedanken experiments, in particular case D, and reviewing Table \ref{table1}, there is the rather promising theoretical result that, with the correct sources and filters, two-particle interferometry can obtain both the mass and the nature of the neutrino of any flavor using a single physical-observable, $C_2$.  While the above experimental discussion is not meant to be exhaustive, it appears the practical requirements currently render the method prohibitive and would require a fundamental shift in the way neutrinos are detected.

\begin{acknowledgments}
The author would like to especially thank N.~Xu for inspiring and invaluable input.  Thanks also to J.~L.~Klay and my other colleagues at LBNL and LLNL for stimulating discussions.   This work was supported by the Director, Office of Science, of the U.S. Department of Energy under Contract No.~DE-AC02-05CH11231.





\end{acknowledgments}

\bibliography{nunuHBT}

\end{document}